\documentstyle[aps,prl,twocolumn]{revtex}

\author{Dmitry V. Savin$^{1,2}$, Yan V. Fyodorov$^{3}$, and Hans-J\"urgen
Sommers$^1$}
\address{$^1$Fachbereich Physik, Universit\"at-GH Essen, 45117 Essen,
Germany}
\address{$^2$Budker Institute of Nuclear Physics,
630090 Novosibirsk, Russia}
\address{$^3$Department of Mathematical Sciences, Brunel University,
Uxbridge, UB8 3PH, United Kingdom}
\title{Reducing nonideal to ideal coupling in random matrix description of  
\\ chaotic scattering: Application to the time-delay problem}
\date{Received 8 November 2000;
published in Phys. Rev. E {\bf 63} (March 2001), Rapid Communication}

\begin{document}
\draft
\maketitle

\begin{abstract}
We write explicitly a transformation of the scattering phases
reducing the problem of
quantum chaotic scattering for systems with $M$ statistically equivalent
channels at nonideal coupling to that for ideal coupling. Unfolding the  
phases by their local density leads to universality of their local  
fluctuations for large $M$. A relation between the partial time delays and
diagonal matrix elements of the
Wigner-Smith matrix is revealed for ideal coupling. This helped
us in deriving the joint probability distribution of partial time
delays and the distribution of the Wigner time delay.
\end{abstract}

The random matrix theory (RMT) is generally accepted to be an
adequate tool for describing various universal
statistical properties of
quantum systems with chaotic intrinsic dynamics, see
Ref.\cite{GMGW98} and references therein. In particular, one can distinguish  
two variants of the RMT approach allowing one to address
the chaotic nature of quantum scattering.
The first one \cite{MPS85} considers the scattering matrix $S$
as the prime object without any reference to the system Hamiltonian.
The probability distribution
$P(S)$ of $S$ at the fixed energy $E$ of incident particles is chosen
to satisfy a maximum entropy principle and  natural constraints which follow
from the unitarity and causality of $S$, and the presence (or absence) of
the time-reversal (TRS) and spin-rotation (SRS) symmetries,
\begin{equation} \label{Poisson}
P(S) \propto \left| \frac{\det(1- \bar{S}^{\dag}\bar{S})}{
\det(1- \bar{S}^{\dag}S)^2}\right|^{(\beta M +2-\beta)/2} \,.
\end{equation}
Such a distribution is known as the Poisson kernel \cite{Hua} and uses the
phenomenological average (or optical) $S$-matrix $\overline{S(E)}$ as
the set of input parameters. Without loss of generality $\bar{S}$ can be  
considered as diagonal \cite{EW73}.
$P(S)$ depends also on the
number of scattering channels $M$ and the symmetry index $\beta$
[$\beta$=2 for a system with broken TRS,  and $\beta$=1(4) if the TRS is  
preserved and the SRS is present (absent)].

The approach proved to be a success for extracting many characteristics
important in the theory of mesoscopic transport\cite{BeenRev}. However,
correlation properties of the $S$-matrix at close values of energy $E$
as well as spectral characteristics of an open system
related to the so-called resonances turn out to be inaccessible in the
framework of such an approach, essentially
because of the one-energy nature of the latter.
To address such quantities one needs to consider the Hamiltonian
$\hat{H}$ of the quantum chaotic system as the prime building block of the  
theory. It amounts to treating $\hat{H}$ as a
large $N\times N$ random matrix of appropriate symmetry
and relating $S$ to the Hamiltonian by means
of standard tools of the scattering theory\cite{VWZ85,SZ89}.
This idea supplemented with the
supersymmetry technique of ensemble averaging \cite{Efetov}
resulted in advance in calculating $S$-matrix correlation
functions\cite{VWZ85,LSSSa} and many other related characteristics such as,  
e.g., time delays \cite{LSSSb,FS97,FSS97}, see Refs.\cite{FS97,GMGW98} for
a review.

In the limit $N$$\to$$\infty$ one can prove\cite{B95}  the equivalence of  
both mentioned approaches  by deriving the Poisson kernel (\ref{Poisson})  
from the Hamiltonian approach (see also Ref.\cite{FS97}),  with the
average $S$-matrix being
\begin{equation} \label{Sav}
\overline{S(E)}_{ab} = \frac{1-\gamma_a[iE/2+\pi\nu(E)]}{
1+\gamma_a[iE/2+\pi\nu(E)]}\,\delta_{ab}\,,
\end{equation}
independent of $\beta$.
Here, the average density of states
$\nu(E)\!=\!\pi^{-1}\sqrt{1\!-\!(E/2)^2}$ determines the mean level spacing
$\Delta$=$(\nu N)^{-1}$ of the closed system, and phenomenological constants
$\gamma_c\!>\!0$  characterize the coupling strength
to continuum in different
scattering channels ($c$=1,\ldots,$M$).

The particular case of ideal coupling, $\bar{S}$=0 [when the transmission  
coefficients equal unity for all channels, see Eq.~(\ref{T}) below],
plays an especiallly important role for the $S$-matrix approach  
\cite{noteS}. Equation (\ref{Poisson}) simplifies then to $P_0(S)$=const,  
which is invariant under the transformations of $S$,
leaving the measure invariant. Such a situation
corresponds to the so-called Dyson's circular ensemble
(CE) of unitary matrices and is much
simpler to handle analytically.

A general situation of nonideal coupling,  $\bar{S}\!\neq\!0$,
turns out to be much more complicated.
It is natural to expect, however,
that results obtained for the case of nonideal coupling
could be related to those at ideal coupling. Although many
useful ideas around such a relation were discussed
in the literature \cite{MPS85,B95,FS97,GM98}, we are not aware of explicit  
relations,  to the best of our knowledge.

In this Rapid Communication
we consider the most simple but physically important case of
statistically equivalent channels. We demonstrate the
validity  of the following simple
statement (and discuss several applications of it):
Let $S(E)\!=\!U\hat{s}(E)U^{\dag}$, where
$\hat{s}(E)\!=\!\mbox{diag}(e^{2i\delta_1(E)},\ldots,e^{2i\delta_M(E)})$,
be the random $S$-matrix at the energy $E$, the distribution of which
is given by the Poisson kernel (\ref{Poisson}) with an explicit
parameterization of $\overline{S(E)}$ from Eq.~(\ref{Sav})
($\gamma_c$=$\gamma$ for all $c$). Then for every $E$ the transformation of
the eigenphases $\delta_c(E)$
\begin{equation} \label{map}
\phi_c = \arctan\left[ \frac{1}{\pi\nu(E)} \left( \frac{1}{\gamma}
         \tan \delta_c(E) + \frac{E}{2}\right)\right]
\end{equation}
maps them to the eigenphases $\phi_c$ of the random scattering matrix, the
distribution of which is given by the CE of the same symmetry.
In particular, the joint probability density function (JPDF) of
$\phi_c$ is \cite{GMGW98}
\begin{equation} \label{phi}
p_0\bigl(\{\phi_c\}\bigr) \propto \prod_{a<b}
\left|e^{2i\phi_a}-e^{2i\phi_b}\right|^{\beta}\,.
\end{equation}
The matrix $U$ of energy dependent eigenvectors uniformly distributed in the  
orthogonal,  unitary, or symplectic group (for
$\beta$=1,2, or 4, respectively) is not affected by map (\ref{map}).
This is a consequence of statistical
equivalence of the scattering channels.

The suggested transformation was first noticed and verified
in Ref.\cite{FS97} for
the case of broken TRS ($\beta$=2) and further exploited in
Ref.\cite{FSS97}. It can be easily generalized to the
other symmetry classes as follows.
We calculate first the Jacobian of the transformation
(\ref{map}). After a simple algebra it can be represented as
\begin{equation}\label{Jac}
\frac{\partial\phi_c}{\partial\delta_c} =
\frac{\cos^2\phi_c}{\pi\nu\gamma\cos^2\delta_c} =
\frac{T}{|1-\overline{S}^{*} e^{2i\delta_c}|^2}\,.
\end{equation}
where
\begin{equation}\label{T}
T(E) = 1-|\overline{S(E)}|^2
     = \frac{4\gamma\,\pi\nu(E)}{1+\gamma^2+2\gamma\,\pi\nu(E)}\,.
\end{equation}
is the energy dependent transmission coefficient\cite{VWZ85}.
We note that exactly the same factors as Eq.~(\ref{Jac}) appear in  
Eq.~(\ref{Poisson}). Employing the identity
\begin{equation}\label{}
\left|e^{2i\phi_a}-e^{2i\phi_b}\right| =
\left|e^{2i\delta_a}-e^{2i\delta_b}\right|
\left|\frac{\cos\phi_a\cos\phi_b}
{\pi\nu\gamma\cos\delta_a\cos\delta_b}\right|\,,
\end{equation}
we substitute Jacobians (\ref{Jac})
into Eq.~(\ref{phi}) and, making use of the identity  
$\prod^{M}_{a<b}f_af_b=(\prod_{c}f_c)^{M-1}$, arrive at
\begin{equation}\label{P-phases}
p\bigl(\{\delta_c\}\bigr) \propto \prod_{a<b}
\left|e^{2i\delta_a}-e^{2i\delta_b}\right|^{\beta}
\prod^{M}_{c=1}\left|\frac{\partial\phi_c}{\partial\delta_c}
\right|^{\beta(M-1)/2+1}.
\end{equation}
With Eq.~(\ref{Jac}) taken into account, we immediately recognize in this
expression the JPDF  of the eigenphases corresponding to Poisson kernel
(\ref{Poisson}). Due to the scalar nature of transformation (\ref{map})
it does not change the matrix $U$ of eigenvectors.

Let us start with considering the mean density $\rho(\delta)$ of
scattering (eigen)phases at arbitrary coupling. It is self-evident
that the phases in the CE
(i.e. for the case $\overline{S}$=0) are uniformly distributed on the unit
circle, the average density being merely
$\rho_0(\phi)\!=\!(1/M)\sum_c\,\overline{\delta(\phi\!-\!\phi_c)}\!=\!1/\pi$.
The corresponding density for $\overline{S}\!\neq\!0$ is not constant.
Indeed,  using the identity
$\rho(\delta)d\delta\!=\!\rho_0(\phi)d\phi$,
we see that
\begin{equation} \label{rho}
\rho(\delta) = \frac{1}{\pi}\left|\frac{\partial\phi}{\partial\delta}\right|
= \frac{T}{|1-\overline{S}^{*} e^{2i\delta}|^2} \,.
\end{equation}
Although simple, this relation is an important one
and establishes the physical meaning of the Jacobians of
transformation (\ref{map}) relating them to the
corresponding densities of the scattering phases. Density (\ref{rho}), being  
expressed in terms of $\overline{S}$ only, does not depend on the particular  
choice of $\overline{S}$ used in the derivation as long as the average  
$S$-matrix is proportional to the unit matrix.

It is instructive to look at Eq.~(\ref{P-phases}) in the limit of large
number of channels when the typical difference
$\delta_a\!-\!\delta_b\! \sim\! 1/M \!\ll\!1$. Then one can expand  
$\delta_c\!=\!\delta_0 \!+\! \tilde\delta_c$
($\tilde\delta_c$$\ll$1) around, say, $\delta_0$. The leading
contribution  is given by
$p\bigl(\{\tilde\delta_c\}\bigr)$$\propto$ $\prod_{a<b}
|\tilde\delta_a\!-\!\tilde\delta_b|^{\beta} \prod_{c}
|\partial\phi_c/\partial\delta_c|^{\beta(M-1)/2+1}_{\delta_0}$,
which further goes to
$ p_0\bigl(\{\tilde\phi_c\}\bigr)$$\propto$$ \prod_{a<b}
|\tilde\phi_a\!-\!\tilde\phi_b|^{\beta}$
and agrees with distribution (\ref{phi}) of the CE
upon the proper rescaling of the phases,
\begin{equation} \label{incr}
\tilde\phi_c = \left|\partial\phi_c/\partial\delta_c\right|_{\delta_0}
\tilde\delta_c = \pi\rho(\delta_0)\tilde\delta_c\,.
\end{equation}

We see that in the limit $M\gg 1$
the local fluctuations of the phases unfolded by their local density
turn out to be uniformly described by
the CE at arbitrary coupling strength.
Such a universality in statistics of
phases of random unitary (scattering) matrices has
much in common with that typical for
eigenvalues of random Hamiltonian matrices \cite{GMGW98} and is in agreement  
with results of realistic numerical simulations for $M=23$\cite{DLS96}.

Let us now consider an application of the same
ideas to the time-delay problem, where such a
universality reveals itself explicitly. Following
the original wave-packet analysis by Eisenbud, Wigner and Smith
\cite{W55}  it is natural to define \cite{FS97} the {\it partial} time
delays via the energy derivative of the scattering phases,
$\tau_c=2\hbar\partial\delta_c/\partial E$. Their statistical properties
have been studied in much detail in the framework of the Hamiltonian
approach for the case of broken \cite{FS97} and preserved TRS as well as in
the whole crossover region of gradually broken TRS \cite{FSS97}.
Recently, some of these predictions were successfully verified on the model
of a quantum Bloch particle chaotically moving in a superposition of
ac and dc fields\cite{Kol}.

In particular, the mean density of partial time delays
${\cal P}(\tau) \!=\! (1/M)\sum_c\,\overline{\delta(\tau \!-\! \tau_c)}$
turns out to be especially simple at ideal coupling, $T$=1,
when it reads as
\begin{equation} \label{P_0}
{\cal P}_0(t \!=\! \tau/t_H) = \frac{(\beta/2)^{\beta M/2}}{\Gamma(\beta M/2)}
\frac{e^{-\beta/2t} }{ t^{\beta M/2+2} }  \,,
\end{equation}
with $t_H$=$2\pi\hbar/\Delta$ being the Heisenberg time.
Due to Eq.~(\ref{incr}), the
partial time delays at ideal and nonideal coupling ($\tau^{(0)}_c$ and
$\tau_c$, respectively) are simply related as
\begin{equation} \label{tau_0tau}
\tau^{(0)}_c = 2\hbar \partial\phi_c/\partial E = \pi\rho(\delta_c)\tau_c
\,.
\end{equation}
Here, we have neglected the smooth nonresonant dependence of
$\rho(\delta)$ on $E$.
Since the phase and its derivative (the partial time delay) are uncorrelated
quantities in the CE \cite{BFB97}, their joint distribution factorizes:
$\widehat p_{0}(\phi,\tau^{(0)}) \!=\! (1/\pi){\cal P}_0(\tau^{(0)})$.
This is not the case for $\widehat p(\delta,\tau)$, when
$\overline{S} \!\neq\! 0$. The relation
$\widehat p_{0}(\phi,\tau^{(0)})d\phi\,d\tau^{(0)} \!=\!
\widehat p(\delta,\tau)d\delta\,d\tau$ between them allows us, however, to
represent the density of partial time delays at nonideal
coupling as
\begin{eqnarray} \label{P}
{\cal P}(\tau) &=& \int^{\pi}_{0}\!\!d\delta
\left| \frac{ \partial(\phi,\tau^{(0)}) }{ \partial(\delta,\tau) }\right|
\widehat p_0\bigl(\phi(\delta),\tau^{(0)}(\delta,\tau)\bigr) \nonumber \\
&=& \int^{\pi}_{0}\!\!\frac{d\delta}{\pi}\, [\pi\rho(\delta)]^2\,
{\cal P}_0(\pi\rho(\delta)\tau) \,.
\end{eqnarray}
One can easily convince oneself \cite{comment1} that such a
formula reproduces in every detail the expression obtained in
Ref.\cite{FS97} by means of supersymmetry calculations.
It is worth mentioning that the density of
phases (\ref{rho}) is independent of the underlying symmetry and
therefore  Eq.~(\ref{P}) is also valid
for the crossover regime of partly broken TRS.
[Note that in the crossover regime ${\cal P}_0(t)$ is a slightly more  
complicated function, see Ref.\cite{FSS97}].

Expression (\ref{P}) is the proper one for generalization to the JPDF of
the partial time delays, $w\bigl(\{\tau_c\}\bigr)$. Before doing this, we
first establish a useful relation between $\tau_c$ and the matrix elements
of the Wigner-Smith time-delay matrix  $Q=-i\hbar(\partial S/\partial
E)S^{\dag}$ \cite{W55}. Writing $S$ in the eigenbasis representation as
$S=U\hat{s}U^{\dag}$, one obtains
\begin{equation} \label{Q}
U^{\dag}QU = -i\hbar\frac{\partial\hat{s}}{\partial E}\hat{s}^{\dag}
+ i\hbar\left[\hat{s}, U^{\dag}\frac{\partial U}{\partial E}\right]
\hat{s}^{\dag}  \,,
\end{equation}
where $[,]$ denotes the commutator.  The matrix $\hat{s}$ being
diagonal, the diagonal elements of the second term in Eq.~(\ref{Q}) are zero,
whereas the first term  is exactly the diagonal matrix of the partial time
delays. Thus, the partial time delays coincide with the diagonal elements of
the time-delay matrix taken in the eigenbasis of the scattering matrix,
\begin{equation} \label{tau}
\tau_c = [U^{\dag}QU]_{cc} \,.
\end{equation}

The physical meaning of the diagonal elements of the time-delay
matrix is well known: they
describe the time delay of a wave-packet incident in a given
channel \cite{W55,L77}. Thus, relation
(\ref{Q}) sheds more light on the physical meaning of the
somewhat formally defined partial time delays.
In particular, one expects that for the case of ideal coupling the
inherent rotational invariance of the problem makes all the
basises statistically equivalent and thus the JPDF
of diagonal elements of the $Q$-matrix should coincide with
that of partial time delays.

The latter claim can be substantiated as follows.
Following the insightful paper \cite{BFB97},
it is convenient to consider the ``symmetrized''
time-delay matrix $Q_{\mbox{\scriptsize s}}$ \begin{equation} \label{Qs}
Q_{\mbox{\scriptsize s}} = S^{-1/2}QS^{1/2} = -i\hbar
S^{-1/2}\frac{\partial S}{\partial E}S^{-1/2}\,.
\end{equation}
This similarity transformation unveils the symmetry which is hidden in $Q$:
$Q_{\mbox{\scriptsize s}}$ is already a real symmetric (hermitian, or
quaternion self-dual) matrix for $\beta=1$, (2, or 4). In the eigenbasis of
$S$ the diagonal elements of $Q_{\mbox{\scriptsize s}}$ and those of $Q$
coincide. Moreover, in the case of chaotic scattering with ideal coupling,
the matrix $Q_{\mbox{\scriptsize s}}$ turns out to
be statistically independent of $S$, their joint probability density being
$\widehat P_0(S,Q_{\mbox{\scriptsize s}}) \!=\!
P_0(S)W_0(Q_{\mbox{\scriptsize s}})$, where
\begin{equation} \label{QS}
W_0(Q_{\mbox{\scriptsize s}})  \propto \theta(Q_{\mbox{\scriptsize s}})
\det(Q_{\mbox{\scriptsize s}})^{-3\beta M/2-2+\beta}
e^{-(\beta/2) t_H \mbox{\scriptsize tr}Q^{-1}_{\mbox{\scriptsize s}}}
\end{equation}
is the probability density of the time-delay matrix \cite{BFB97}.
The latter is manifestly invariant under the choice of the basis for
$Q_{\mbox{\scriptsize s}}$ proving the above statement on the relation
between statistics of
partial time delays and diagonal elements of the Wigner-Smith matrix.

To find the corresponding JPDF $w_0\bigl(\{\tau_c\}\bigr)$
one has to integrate out all off-diagonal elements of
$Q_{\mbox{\scriptsize s}}$ which is a hard problem in general.
For the case of unitary symmetry, $\beta=2$, one can perform the job
by splitting the integration into that over the matrix
$\hat{q}=\mbox{diag}(q_1,...,q_M)$ of eigenvalues of $Q_{\mbox{\scriptsize
s}}$ and that of the eigenvectors, $V$,
\begin{equation}\label{unit}
w^{u}_0\bigl(\{\tau_c\}\bigr)  \propto  \int\!\! d[\hat{q}]\,
\frac{\theta(\hat{q})\Delta^2(\hat{q})}{\det(\hat{q})^{3M}}
e^{-t_H\mbox{\scriptsize tr}\hat{q}^{-1}}
{\cal Q}\bigl(\{\tau_c\}\bigr)\,,
\end{equation}
with $\Delta(\hat{q})=\prod_{a<b}(q_a-q_b)$ being the Vandermonde
determinant. Here
${\cal Q}\bigl(\{\tau_c\}\bigr)  =  \int\!\! d[V]
\prod_{c=1}^M\delta\bigl(\tau_c-(V\hat{q}V^{\dagger})_{cc} \bigr)$
stands for the remaining integral over the unitary group which can be done,   
following Ref.\cite{FK99}, by means of
the famous Itzykson-Zuber formula\cite{IZ}.
Finally, we find it more convenient to
define the generating function of partial time delays rather than
the JPDF itself and obtain
\begin{equation} \label{moments}
\left\langle e^{-i(k_1\tau_1+\dots+k_M\tau_M)} \right\rangle_{\tau} \propto
\frac{\det \left[\psi_j(k_l)\right]}{\prod_{a<b}(k_a-k_b)}\,,
\end{equation}
where
$\psi_j(k_l)=\int_0^{\infty}\!dq\,{q^{j-3M}} e^{-ik_l q-t_H/q}$,
the index $l$ spans the values $l=1,\ldots,M$ and
$j=0,1,\ldots,M-1$.

Such an expression allows us to calculate all the moments and
correlation functions of partial time delays by a simple differentiation.
Moreover, setting in the preceding equation $k_1$=\ldots=$k_M$=$k$
and calculating the corresponding limit in the right-hand side, we come
to a convenient representation for the distribution ${\cal P}^{u}_M(t_w)$
of the Wigner time delay,
$t_w \!=\! (\tau_1$+...+$\tau_M)/Mt_H$,
for a system with broken TRS and ideal coupling to continuum,
\begin{equation}\label{Wigner}
{\cal P}^{u}_M(t_w)\propto\int^{\infty}_{-\infty}\!\!dk e^{iMkt_w}
\det\bigl[\psi^{(n)}_j(k)\bigr]\,,
\end{equation}
where $\psi^{(n)}_j(k) \!\equiv\! d^n\psi_j(k)/dk^n$,
and $j,n \!=\!0,\dots,M\!-\!1$.

The distribution of the Wigner time delay was earlier calculated explicitly  
only for the case of $M$=1\cite{FS97,FSS97,GMB96}, when it follows from
Eq.~(\ref{P_0}). Compact expression (\ref{Wigner}) is valid for $\beta$=2  
and arbitrary $M$\cite{comment2}. For $M$=2, Eq.~(\ref{Wigner}) can be  
integrated further to yield
\begin{equation}\label{M=2}
{\cal P}_2(t_w) \propto t_w^{-3(\beta+1)}e^{-\beta/t_w}
U\bigl({\textstyle \frac{\beta+1}{2},2\beta+2,\frac{\beta}{t_w} }\bigr) \,,
\end{equation}
with
$U(a,b,z)=[1/\Gamma(a)]\int^{\infty}_{0}\!\!\!dy\,y^{a-1}(1+y)^{b-a-1}e^{-zy}$
being the confluent hypergeometrical function.
Here we represented the above distribution (\ref{M=2}) in a form covering
all $\beta=1,2,4$ which will be verified below.
In particular, the asymptotic behavior at $t_w\gg 1$ is
${\cal P}_2(t_w)\propto t_w^{-\beta-2}$ in
agreement with the known universal tail $t^{-\beta M/2-2}$, which is typical
for the time-delay distributions in open chaotic systems
\cite{FS97,FSS97,BFB97,Kol}.

To verify Eq.~(\ref{M=2}) for $\beta$=1,4,
it is convenient to consider a general problem of finding the
distribution
$\widetilde W_0(\widetilde Q)$ of the $n\times n$ submat\-rix $\widetilde Q$
standing on the main diagonal of $Q_{\mbox{\scriptsize s}}$. This
distribution is found to be
\begin{equation} \label{Qsub}
\widetilde W_0(\widetilde Q)  \!\propto\! \theta(\widetilde Q)
\det(\widetilde Q)^{-\beta(M/2+n-1)-2}
e^{-\beta t_H \mbox{\scriptsize tr} {\widetilde Q}^{-1}/2}.
\end{equation}
The particular case $n$=1 reproduces result (\ref{P_0}) of the Hamiltonian  
approach.
Equation (\ref{Qsub}) for $n$=2 helps in calculating the joint distribution
$\widehat w_0(t_1,t_2)$ of two partial time delays
$t_{1,2}\!=\!\tau_{1,2}/t_H$ for arbitrary $M$. One obtains
\begin{equation} \label{two}
\frac{\widehat w_0(t_1,t_2)}{{\cal P}_0(t_1){\cal P}_0(t_2)} \propto
\frac{ U\bigl({\frac{\beta}{2}, \frac{\beta M}{2}\!+\!\beta\!+\!2,
\frac{\beta}{2t_1}\!+\! \frac{\beta}{2t_2} }\bigr) }{ (t_1t_2)^{\beta/2} }
\,.
\end{equation}
The knowledge of $\widehat w_0(t_1,t_2)$ allows us to find further
the distribution of the Wigner time delay for $M$=2 and thus prove
the formula (\ref{M=2}) for any $\beta$.
As follows also from Eq.(\ref{two}), there exist nonvanishing
correlations between the partial time delays. They are, however, of  
different nature as compared to the correlations between the {\it proper}  
time delays
(the eigenvalues of $Q$) which show repulsion \cite{BFB97}.


For $\bar{S}\!\neq\!0$ the matrices $S$ and $Q_{\mbox{\scriptsize s}}$ cease  
to be statistically independent variables and do correlate.
Therefore statistical properties  of diagonal elements of $Q$
in arbitrary basis (save the eigenbasis of $S$) are different from that of
partial time delays, unless coupling is ideal. Still, the
JPDF $w\bigl(\{\tau_c\}\bigr)$ of the partial time delays at
nonideal coupling can be found by repeating
basically the same steps which lead to Eq.~(\ref{P}). The identity
$\widehat p\bigl(\{\delta_c\},\{\tau_c\}\bigr)d[\delta]\,d[\tau] =
\widehat p_0\bigl(\{\phi_c\},\{\tau^{(0)}_c\}\bigr)d[\phi]\,d[\tau^{(0)}]$,
together with the statistical independence of $\phi_c$ and $\tau^{(0)}_c$
[which follows from Eq.~(\ref{QS})], allows us to relate
$w\bigl(\{\tau_c\}\bigr)$ and  $w_0\bigl(\{\tau_c\}\bigr)$ as follows:
\begin{eqnarray} \label{JDPF}
w(\{\tau_c\}) &=& \int\!\!d[\delta]
\left| \frac{ \partial\bigl(\{\phi_c\},\{\tau^{(0)}_c\}\bigr) }{
\partial\bigl(\{\delta_c\},\{\tau_c\}\bigr) }\right|
\widehat p_0\bigl(\{\phi_c\},\{\tau^{(0)}_c\}\bigr) \nonumber \\
&=& \int\!\!d[\delta]\,\prod^{M}_{c=1} [\pi\rho(\delta_c)]\,
p\bigl(\{\delta_c\}\bigr) w_0\bigl(\{\pi\rho(\delta_c)\tau_c\}\bigr) \,,
\end{eqnarray}
where $d[\delta]$ means the product of differentials.

In conclusion, we suggest the transformation of the scattering phases,  
allowing one to reduce the problem of quantum chaotic scattering  with  
statistically equivalent channels at arbitrary coupling to that for ideal  
coupling. Applications of this transformation to statistical properties of  
phases and those of time delays are discussed.

We are grateful to V.V. Sokolov for critical comments.
The financial support by
SFB 237 ``Unordnung and grosse Fluktuationen'' (D.V.S. and H.J.S.),
RFBR Grant No. 99--02-16726 (D.V.S.), and EPSRC Grant No. GR/R13838/01
(Y.V.F.) is acknowledged with thanks.



\begin{thebibliography}{99}
\vspace{-1.5 cm}
\bibitem{GMGW98}
T. Guhr {\it et al.},
Phys. Rep. {\bf 299},
189 (1998).
\bibitem{MPS85}
P.A. Mello , P. Pereyra, and T.H. Seligman, Ann. Phys. {\bf 161}, 254
(1985);
W.A. Friedman and P.A. Mello, Ann. Phys. {\bf 161}, 276 (1985).
\bibitem{Hua}
L.K. Hua, {\it Harmonic Analysis of Functions of Several Complex Variables
in the Classical Domains} (AMS, Providence, 1963).
\bibitem{EW73}
C.A. Engelbrecht and H.A. Weidenm\"uller, Phys. Rev. C {\bf 8}, 859 (1973).
\bibitem{BeenRev}
C.W.J. Beenakker, Rev. Mod. Phys. {\bf 69}, 731 (1997).
\bibitem{VWZ85}
J.J.M. Verbaarschot {\it et al.},
Phys. Rep. {\bf 129}, 367 (1985).
\bibitem{SZ89}
V.V. Sokolov and V.G.  Zelevinsky, Phys. Lett. B {\bf 202},
  140 (1988); Nucl. Phys. A {\bf 504}, 562 (1989).
\bibitem{Efetov}
K.B. Efetov, Adv. Phys. {\bf 32}, 53 (1983).
\bibitem{LSSSa}
N. Lehmann  {\it et. al.}, Nucl. Phys. A {\bf 582}, 223 (1995).
\bibitem{LSSSb}
N. Lehmann {\it et. al.}, Physica D {\bf 86}, 572 (1995).
\bibitem{FS97}
Y.V. Fyodorov and H.-J. Sommers, J. Math. Phys. {\bf 38}, 1918 (1997)
\bibitem{FSS97}
Y.V. Fyodorov,  D.V. Savin, and H.-J. Sommers,
Phys. Rev. E {\bf 55}, R4857 (1997).
\bibitem{B95}
P.W. Brouwer, Phys. Rev. B {\bf 51}, 16878 (1995).
\bibitem{noteS}
It is worth stressing that $\bar{S}$=0 requires
both $E$=0 and all $\gamma_c$=1, according to Eq.~(\ref{Sav}).
\bibitem{GM98}
V.A. Gopar and P.A. Mello, Europhys. Lett. {\bf 42}, 131 (1998).
\bibitem{DLS96}
B. Dietz, M. Lombardi, and T.H. Seligman, Phys. Lett. A {\bf 215}, 181 (1996).
\bibitem{W55}
E.P. Wigner, Phys. Rev. {\bf 98}, 145 (1955);
F.T. Smith, Phys. Rev. {\bf 118}, 349 (1960).
\bibitem{Kol} M. Gl\"{u}ck, A.R. Kolovsky, and H.-J. Korsch,
Phys. Rev. E {\bf 60}, 247 (1999).
\bibitem{BFB97}
P.W. Brouwer, K.M. Frahm, and C.W.J. Beenakker, Phys. Rev. Lett. {\bf 78},
4737 (1997); Waves in Random Media {\bf 9}, 91 (1999).
\bibitem{comment1}
For convenience of the reader we note that the identity
$\int^{2\pi}_0\!\!d\delta\,f(p\cos\delta\!+\!q\sin\delta) \!=\!
2\int^{\pi}_0\!\!d\delta\,f(\sqrt{p^2\!+\!q^2}\cos\delta)$
allows us to write Eq.(\ref{P}) in the form of Refs.\cite{FS97,FSS97},
with $\pi\rho(\delta)$
being replaced by
$\pi\widetilde\rho(\delta)\!=\![g-\sqrt{g^2\!-\!1}\cos\delta]^{-1}$,
$g\!=\!2/T\!-\!1$.
\bibitem{L77}
V.L. Lyuboshitz, Phys. Lett.  B {\bf 72}, 41 (1977);
Sov. J.  Nucl. Phys. {\bf 27}, 502 (1978); JETP Lett. {\bf 28}, 30 (1978).
\bibitem{FK99}
Y.V. Fyodorov and B.A. Khoruzhenko, Phys. Rev. Lett. {\bf 83}, 65 (1999).
\bibitem{IZ} C. Itzykson and J.B. Zuber, J.Math.Phys. {\bf 21}, 411 (1980).
\bibitem{GMB96}
V.A. Gopar, P.A. Mello, and M. B\"uttiker, Phys. Rev. Lett. {\bf 77}, 3005
(1996).
\bibitem{comment2}
We note, that due to the relation $t_w$=tr$Q_{\mbox{\scriptsize s}}/Mt_H$,  
Eq.~(\ref{Wigner}) can be also derived directly from Eq.~(\ref{QS}).
\end{thebibliography}
\end{document}